\documentclass[twocolumn,aps,superscriptaddress,showpacs,nofootinbib,floatfix]{revtex4}
\usepackage{float}
\usepackage{epsfig,bm,feynmf}
\usepackage{graphics}

\usepackage[normalem]{ulem}  
\usepackage[dvips]{color} 

\renewcommand\sout{\bgroup \color{red} \ULdepth=-.5ex \ULset}

\begin{document}
	
	\title{Spinodal Instabilities of Baryon-Rich Quark-gluon Plasma in the PNJL Model}
	\author{Feng Li}\email{lifengphysics@gamil.com}
	\affiliation{Cyclotron Institute and Department of Physics and
		Astronomy, Texas A$\&$M University, College Station, TX 77843-3366,
		USA}
	\author{Che Ming Ko}\email{ko@comp.tamu.edu}
	\affiliation{Cyclotron Institute and Department of Physics and
		Astronomy, Texas A$\&$M University, College Station, TX 77843-3366, USA}
	
\begin{abstract}
Using the Polyakov-Nambu-Jona-Lasinia (PNJL) model, we study the spinodal instability of a baryon-rich quark-gluon plasma in the linear response theory.  We find that the spinodal unstable region in the temperature and density plane shrinks with increasing wave number of the unstable mode and is also reduced if the effect of Polyakov loop is not included.  In the small wave number or long wavelength limit, the spinodal boundaries in both cases of with and without the Polyakov loop coincide with those determined from the isothermal spinodal instability in the thermodynamic approach.  Also, the vector interactions among quarks is found to suppress unstable modes of all wave numbers. Moreover, the growth rate of unstable modes initially increases with the wave number but is reduced when the wave number becomes large. Including the collisional effect from quark scattering via the linearized Boltzmann equation, we further find that it decreases the growth rate of unstable modes of all wave numbers. Relevance of these results to relativistic heavy ion collisions is discussed.
\end{abstract}

\maketitle

\section{Introduction}

Understanding the phase transition from the quark-gluon plasma (QGP) to the hadronic matter is currently of great  interest. According to lattice QCD calculations, the phase transition is a smooth crossover if the QGP has zero baryon chemical potential~\cite{Aoki:2006we}. However, due to difficulties in treating the fermion sign problem~\cite{PhysRevB.41.9301}, lattice QCD has not provided definitive information on the order of the phase transition in QGP of finite chemical baryon potential. On the other hand, studies based on various theoretical models have indicated that the quark-gluon plasma to hadronic matter transition changes to a first-order one
when its baryon chemical potential is larger than a critical value.  A possible signal for such a critical end point, at which the crossover transition changes to a first-order one, is the large higher moments in the net baryon or proton number event-by-event distributions in heavy ion collisions as suggested in Refs.~\cite{Stephanov:2004wx,Morita:2013tu,Friman:2011zz}.  To determine if the critical point exists and where it is located in the QCD phase diagram, the STAR Collaboration has carried out the beam energy scan (BES) program at RHIC to look for this signal in collisions at energies $\sqrt{s_{NN}}$ ranging from $7.7$ to $39$ GeV~\cite{Nayak:2009wc,Aggarwal:2010wy,McDonald:2015tza}, which is expected to produce a baryon-rich QGP with baryon chemical potential in the range where a first-order QGP to hadronic matter transition is likely to appear. Although no definitive conclusion has been obtained from these experiments on the existence or the location of the critical point, the STAR Collaboration has observed many interesting phenomena that are different from those at higher collision energies. Among them is the increasing splitting of the elliptic flows of particles and their antiparticles with decreasing collision energy~\cite{Adamczyk:2013gv}. Based on an extension of a multiphase transport (AMPT) model~\cite{Lin:2004en} by including  mean-field potentials from the Nambu-Jona-Lasinia (NJL) model~\cite{Nambu:1961tp,Nambu:1961fr} for the partonic phase~\cite{Song:2012cd} and from empirical extracted values for the hadonic phase~\cite{Xu:2012gf}, the authors and their collaborators have obtained a plausible explanation for this experimental observation~\cite{Xu:2013sta}. Since a unique feature of a first-order phase transition is the large density fluctuations due to the spinodal instability that leads to the phase separation, we here extend the above study by using the PNJL model to investigate the spinodal instability of a baryon-rich QGP in the linear response theory as well as its semiclassical approximation, i.e., the linearized Boltzmann equation. We note that there already exist several studies in the literatures on this interesting phenomenon and its possible signals based on the hydrodynamic approach~\cite{Herold201414, Herold:2014uva,PhysRevC.89.034901}.  

The paper is organized as follow. In the next section, we give a brief review on the PNJL model and discuss the thermodynamic properties of a QGP in this model, with particular emphasis on the confinement and mean-field  effects. We then employ in Section III the linear response theory to study the spinodal instability of a baryon-rich quark matter and the growth rate of its unstable modes. The effect of quark scatterings on the growth rate of unstable modes is further studied in  Section IV by using the linearized Boltzmann equation.  Finally, we summarize our study in Section V. 

\section{The PNJL model} 

The PNJL model~\cite{Fukushima:2003fw,Roessner:2006xn} is an extension of the NJL model, which describes the chiral phase transition in a quark matter with the quark condensate $\langle \bar q q\rangle$ as the order parameter, to include the  confinement-deconfinement transition through the Polyakov loop for gluons. In this model, the finite temperature $T=1/\beta$ action of a QGP in the Euclidian time is given by 
\begin{eqnarray}\label{PNJL}
&&\mathcal{S}_\mathrm{PNJL}[q,\bar q, \phi]\nonumber\\
&&=-\int_0^\beta\mathrm d^4 x\bigg\{\bar q(-\gamma^0(\partial_\tau+i\phi)+i\boldsymbol\gamma\cdot\nabla-m)q\nonumber\\
&&+\frac{G_S}{2}\sum_{a=0}^{8}\bigg[(\bar{q}\lambda^aq)^2+(\bar{q}i\gamma_5\lambda^aq)^2\bigg]
-G_V(\bar{q}\gamma_\mu q)^2\nonumber\\
&&-K\bigg[{\rm det}_f\bigg(\bar{q}(1+\gamma_5)q\bigg)+{\rm det}_f\bigg(\bar{q}(1-\gamma_5)q\bigg)\bigg]\bigg\}\nonumber\\
&&+\beta V\mathcal U(\Phi,\bar\Phi,T).
\end{eqnarray}

In Eq.(\ref{PNJL}), the first term with $\phi=0$ is the action from the NJL model based on the Lagrangian for quarks of three flavors~\cite{Bratovic:2012qs}. Specifically, $q=(u,d,s)^T$ is the quark fields, $m={\rm diag}(m_{u}, m_{d}, m_{s})$ are the quark mass matrix, and $\lambda^a$ is the Gell-Mann matrices with $\lambda^0$ being the identity matrix multiplied by $\sqrt{2/3}$. As in the QCD Lagrangian, the NJL Lagrangian preserves the $U(1)\times SU(N_f)_L\times SU(N_f)_R$ symmetry but breaks axial symmetry.  The latter, which is due to the axial anomaly in QCD, is achieved in the NJL model by the Kobayashi-Masakawa-t'Hooft (KMT) interaction~\cite{'tHooft:1976fv} in the last term of its action, where ${\rm det}_f (\bar{q}\Gamma q)=\sum_{i,j,k}\varepsilon_{ijk}(\bar{u}\Gamma q_i)(\bar{d}\Gamma q_j)(\bar{s}\Gamma q_k)$ denotes the determinant in flavor space~\cite{Buballa:2003qv}. Because the NJL model is not renormalizable, a momentum cutoff $\Lambda$ is introduced in all momentum integrations. Taking $\Lambda=0.6023~\mathrm{GeV}$, values of the scalar coupling $G_S$ and the KMT interaction $K$ can be determined from fitting the pion mass, the kaon mass, and the pion decay constant, and their values are $G_S\Lambda^2=3.67$, and $K\Lambda^5=12.36$ if the current quark masses are taken to be $m_{u}=m_d=3.6$ MeV, and $m_s=87$ MeV~\cite{Holstein:1990ua}. For the vector coupling $G_V$, its value affects the order of the quark matter phase transition.  If $G_V$ is large, the first-order phase transition induced by the attractive scalar interaction could disappear. In the present study, we treat it as a parameter to study how it affects the spinodal instability of a baryon-rich quark matter.

The second term in Eq.(\ref{PNJL}) is the action  from the Polyakov loop~\cite{RevModPhys.53.43}, 
\begin{eqnarray}
\Phi({\bf x})\equiv\frac{1}{N_c}{\rm Tr}\left[P\exp\left(i\int_0^\beta d\tau{\phi}({\bf x},\tau)\right)\right],
\end{eqnarray}
through the effective potential $\mathcal U(\Phi,\bar\Phi,T)$,
\begin{eqnarray}
\frac{\mathcal U(\Phi,\bar\Phi,T)}{T^4}&=&-\frac 1 2 a(T)\bar\Phi\Phi+b(T)\ln[1-6\bar\Phi\Phi\nonumber\\&&+4(\bar\Phi^{3}+\Phi^3)-3(\bar\Phi\Phi)^2],
\end{eqnarray}
with
\begin{equation}\label{a}
a(T)=a_0+a_1\left(\frac{T_0} T\right)+a_2\left(\frac{T_0} T\right)^2,\quad b(T)=b_3\left(\frac{T_0} T\right)^3.
\end{equation}
In the above, $N_c=3$ is the number of colors and $\phi$ is a background color field.  In the Polyakov gauge of constant $\phi$ in space and time, the Polyakov loop is simply $\Phi=\frac{1}{3}{\rm Tr}e^{i\phi/T}$ with $\phi={\rm diag}(\phi_1,\phi_2,-\phi_1-\phi_2)$. 

In the mean-field approximation, the constituent quark masses in the PNJL model are determined by the gap equation,
\begin{eqnarray}
M_i&=&m_{i}-2G_S\sigma_i+2K\sigma_j\sigma_k,\quad i=u,d,s
\end{eqnarray}
in terms of the quark condensate
\begin{eqnarray}
\sigma_i&\equiv&\langle\bar q_i q_i\rangle\nonumber\\
&=&6\int_0^\Lambda\frac{\mathrm d^3\mathbf p}{(2\pi)^3} \frac{M_i}{E_{\mathrm p_i}} \left( f_0(\xi_i;\Phi,\bar\Phi) + f_0(\xi_i^\prime;\bar\Phi,\Phi) -1\right).\nonumber\\
\end{eqnarray}
In the above, 
\begin{equation}
f_0(\xi_i;\Phi,\bar\Phi)=\frac{\bar\Phi\xi_i^2+2\Phi\xi_i+1}{\xi_i^3+3\bar\Phi\xi_i^2+3\Phi\xi_i+1}
\end{equation}
is the color averaged distribution of quarks in thermal and chemical equilibrium with $\xi_i=\exp((E_i-\mu+2G_Vn)/T)$ and $\xi^\prime_i=\exp((E_i+\mu-2G_Vn)/T)$, where $E_i=\sqrt{M_i^2+{\bf p}^2}$ and $n$ is the net quark density. It reduces to the normal Fermi-Dirac distribution in the deconfined phase in which $\Phi=\bar\Phi=1$. In the confined phase with $\Phi=\bar\Phi=0$, the color averaged distribution also has a normal Fermi-Dirac distribution but with an effective temperature reduced by a factor of 3 because $f_0(\xi,0,0)=f_0(\xi^3,1,1)$ due to the freeze of color degrees of freedom. As shown later, this would lead to a higher critical temperature in the PNJL model than in the NJL model.

The thermal properties of a QGP of temperture $T$ and volume $V$ can then studied in the PNJL model from the following thermal potential per unit volumn:
\begin{eqnarray}
\label{Omega}
\Omega(T,\mu)&=&\mathcal U(\Phi,\bar\Phi,T)+\sum_iG_S\sigma_i^2-4K\prod_i\sigma_i-G_Vj^2\nonumber\\
&-&\frac T V \ln \mathrm{Det}[\beta\widetilde S^{-1}_{p<\Lambda}]-\frac T V \ln \mathrm{Det}[\beta\widetilde S^{-1}_{p>\Lambda}],
\end{eqnarray}
with $j^\mu=\langle\sum_i\bar q_i\gamma^\mu q_i\rangle$. The $\widetilde S^{-1}_{p<\Lambda}$ and $\widetilde S^{-1}_{p>\Lambda}$ are the Matsubara propagators for quarks with momentum below and above the momentum cutoff $\Lambda$, and they are given by 
\begin{eqnarray}
\ln\det[\beta\widetilde S_{p<\Lambda}^{-1}]&=&2V\sum_{i}\int_0^\Lambda\frac{d^3 \mathbf p}{(2\pi)^3}\bigg\{3\beta E_{i}\nonumber\\
&+&\ln(1+3\bar\Phi\xi_{i}^{-1}+3\Phi\xi_{i}^{-2}+\xi_{i}^{-3})\nonumber\\
&+&\ln(1+3\Phi\xi_{i}^{\prime-1}+3\bar\Phi\xi_{i}^{\prime-2}+\xi_{i}^{\prime-3})\bigg\},\nonumber\\
\ln\det[\beta\widetilde S_{p>\Lambda}^{-1}]&=&2V\sum_{i}\int^\infty_\Lambda\frac{d^3 \mathbf p}{(2\pi)^3}\bigg\{3\beta E^0_{i}\nonumber\\
&+&\ln(1+3\bar\Phi\xi_{0i}^{-1}+3\Phi\xi_{0i}^{-2}+\xi_{0i}^{-3})\nonumber\\
&+&\ln(1+3\Phi\xi_{0i}^{\prime-1}+3\bar\Phi\xi_{0i}^{\prime-2}+\xi_{0i}^{\prime-3})\bigg\},
\end{eqnarray}
where $E_{i}^0=\sqrt{m_{i}^2+\mathbf p^2}$,  $\xi_{0i}=\exp((E^0_{i}-\mu)/T)$ and $\xi^\prime_{0i}=\exp((E^0_{i}+\mu)/T)$. 

Values of $a_i$ in Eq.(\ref{a}) are chosen to reproduce the lattice QCD results for the thermal properties of the pure gauge theory for temperatures up to twice the critical temperature $T_0$ for the confinemnt-deconfinement transition, and this leads to $a_0=3.51$, $a_1=-2.47$, $a_2=15.2$, and $b_3=-1.75$~\cite{deSousa:2010pm}. Depending on the value of $T_0$, the deconfinement transition temperature may or may not coincide with the chiral transition temperature in the PNJL model. For $T_0=270~\mathrm{MeV}$, the deconfinement transition temperature at $\mu=0$ agrees with the chiral transition temperature, and both are higher than that from the lattice QCD calculation. On the other hand, taking $T_0=210~\mathrm{MeV}$ leads to very different temperatures for the deconfinement and chiral transitions, but their average value $T_X=187~\mathrm{MeV}$ is within the range expected from lattice calculations~\cite{Cheng:2009be}.

\begin{figure}[htbp]
	\vspace{-0cm}
	\includegraphics[width=0.4\textwidth]{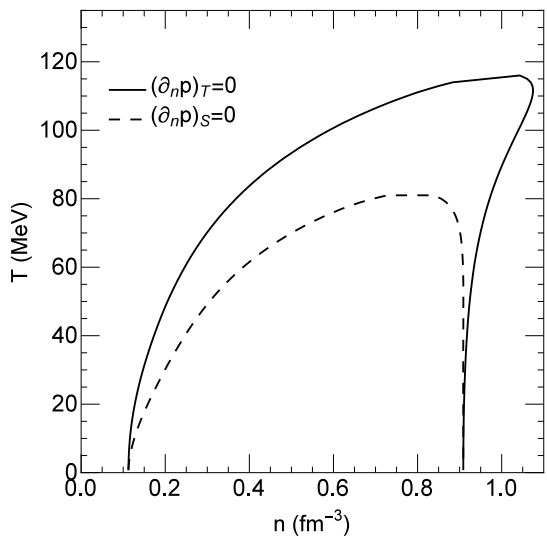}
	\caption{Isothermal ($(\partial_n p)_T=0$) and isentropic ($(\partial_n p)_S=0$) spinodal boundaries from the PNJL model with $G_V=0$.} 	\label{vs0_PNJL}
\end{figure}

\begin{figure}[htbp]
	\vspace{0.5cm}
	\includegraphics[width=0.4\textwidth]{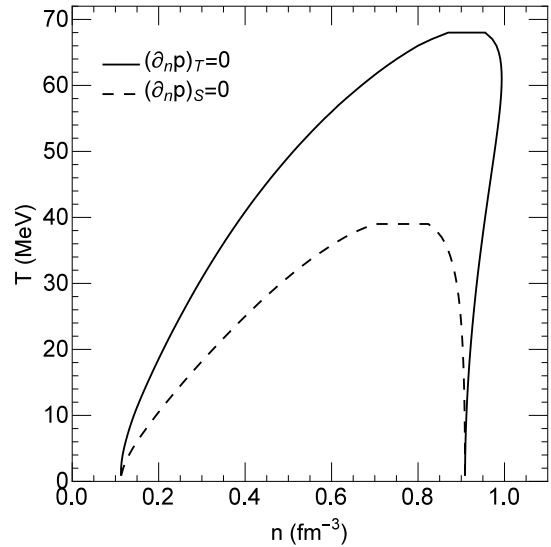}
	\caption{Isothermal ($(\partial_n p)_T=0$) and isentropic ($(\partial_n p)_S=0$) spinodal boundaries from the NJL model with $G_V=0$.} 
	\label{vs0_NJL}
\end{figure}

The spinodal instability in a quark matter occurs when its pressure decreases with increasing density either at constant temperature or constant entropy, i.e., $(\partial_n p)_T<0$ or $(\partial_n p)_S<0$, where the pressure $p$ is related to the thermal potential by $-\Omega(T,\mu)$.  Its boundary, given by $(\partial_n p)_T=0$ (solid line) or $(\partial_n p)_S=0$ (dahed line) in the temperature and density plane, is shown in Fig.~\ref{vs0_PNJL} for the PNJL model with $G_V=0$. It is seen that the isentropic spinodal region is smaller than the isothermal region. Corresponding results without the Polyakov loop, i.e., the NJL model, are shown in Fig.~\ref{vs0_NJL}, and they are similar to those for the PNJL model. However, the critical temperature $T_c$, corresponding to the highest temperature on the isothermal spinodal boundary, in the PNJL model is higher than that in the NJL model, as eluded above on the effect of the Polyakov loop in the confined phase. For a quark matter with temperature and density inside the spinodal region, small density fluctuations will develop into large fluctuations as a result of instability.  In the present study, we study the spinodal instability of quark matter by decomposing it into its Fourier components or unstable modes. 

\section{The linear response theory}

The linear response theory describes the response of an equilibrium system to perturbations. It has been extensively used in calculating the transport coefficients of many-body systems such as the electric conductivity by subject them to an external electric field or the viscosity by applying an external flow field of nonzero gradient.  For studying spinodal instabilities in the linear response theroy, the perturbation is, however, generated internally from the fluctuations that drive the system away from equilibrium.  

\subsection{Theoretical framework}

When a many-body system is slightly perturbed, the deviation of a physical observable from its equilibrium value is proportional to the perturbation.  Such a system can be described by the Hamiltonian: $H=H_0+H^\prime$, where $H_0$ is its Hamiltonian at equilibrium, and $H^\prime$ is the perturbation that drives the system away from equilibrium. In the Schroedinger picture, the density matrix of the system evolves according to
\begin{equation}
 \rho(t)=\mathcal U(t,t_0)\rho(t_0)\mathcal U^\dagger (t,t_0),
\end{equation}
where
\begin{eqnarray}
\mathcal U(t,t_0)\equiv T e^{-i \int_{t_0}^t d\bar t H(\bar t)}
\end{eqnarray}
is the time evolution operator from the initial time $t_0$ to time $t$. Expanding $\mathcal U(t,t_0)$ to the first order in the perturbation $H^\prime$, we have
\begin{eqnarray}
\label{linU}
\mathcal U(t,t_0)&\approx&\mathcal U_0(t,t_0)-i\int_{t_0}^{t}dt^\prime\mathcal U_0(t,t^\prime) H^\prime(t^\prime)\mathcal U_0(t^\prime,t_0),\nonumber\\
\end{eqnarray}
where
\begin{eqnarray}
\mathcal U_0(t,t_0)\equiv T e^{-i \int_{t_0}^t d\bar t H_0(\bar t)}
\end{eqnarray}
is the time evolution operator in the absence of perturbations, the density matrix of the system at time $t$ can then be written as
\begin{eqnarray}
\label{rmtrx}
 \rho(t)\approx \rho_0(t)-i \int_{t_0}^{t}dt^\prime\mathcal U_0(t,t_0)[H_I^\prime(t^\prime),\rho(t_0)]\mathcal U^\dagger_0(t,t_0),\nonumber\\
\end{eqnarray}
where
\begin{equation}
\rho_0(t)=\mathcal U_0(t,t_0)\rho(t_0)\mathcal U^\dagger_0(t,t_0)
\end{equation}
is the density matrix of the system at time $t$ in the absence of perturbations, and
\begin{equation}
H^\prime_I(t^\prime)=U^\dagger_0(t^\prime,t_0)H^\prime(t^\prime)\mathcal U_0(t^\prime,t_0)
\end{equation}
is the perturbation expressed in the interaction picture. 

For a physical observable $A$, its time dependence to the first order in $H^\prime$ is then given by
\begin{eqnarray}
\label{At}
&&\langle A(t)\rangle=\mathrm{tr}[\rho(t)A_S]\approx\langle A_0(t)\rangle\nonumber\\
&+&i \int_{t_0}^{t}dt^\prime \mathrm{tr}\left[\rho(t_0)[H_I^\prime(t^\prime),A_I(t)]\right],
\end{eqnarray}
where the subscription $S$ denotes the Schroedinger picture. The second line in Eq. (\ref{At}) is obtained by employing the cyclic property of the trace operator. For a system that is initially in equilibrium, namely, $\rho(t_0)=\rho_0$, the deviation of $\langle A(t)\rangle$
from $\langle A_0(t)\rangle$ is then 
\begin{equation}
\label{linres}
\delta\langle A(t)\rangle\approx i\int d\bar t \theta(t-\bar t)\langle[H_I^\prime(\bar t),A_I(t)]\rangle_0,
\end{equation}
where the right hand side is the retarded correlator evaluated with the system in an equilibrium state.  Although the correlator generally depends on both $t$ and $\bar t$, it can be simplified to a function of $t-\bar t$ for an equilibrium system.

Since the spinodal instabilities are self-induced, the perturbation $H^\prime_I$ in the PNJL model is just the fluctuations of mean fields, that is
\begin{eqnarray}
H^\prime_I&=&\int d^3 \mathbf x\left[\bar u u\delta M_u+\bar d d\delta M_d+\bar s s\delta M_s\right.\nonumber\\&&\left.+2G_V\delta j_\mu(\bar u\gamma^\mu u+\bar d\gamma^\mu d+\bar s\gamma^\mu s)\right ],
\end{eqnarray}
where the mass fluctuations are given by
\begin{eqnarray}
\delta M_u=-2G_S\delta\sigma_u-2K_S\sigma_s\delta\sigma_d-2K_S\sigma_d\delta\sigma_s,\nonumber\\
\delta M_d=-2G_S\delta\sigma_d-2K_S\sigma_s\delta\sigma_u-2K_S\sigma_u\delta\sigma_s,\nonumber\\
\delta M_s=-2G_S\delta\sigma_s-2K_S\sigma_u\delta\sigma_d-2K_S\sigma_d\delta\sigma_u
\end{eqnarray}
in terms of the fluctuations $\delta\sigma_u$, $\delta\sigma_d$, and $\delta\sigma_s$ of condensates $\sigma_u=\langle\bar uu\rangle$, $\sigma_d=\langle\bar dd\rangle$, and $\sigma_s=\langle\bar ss\rangle$ as well as the fluctuation $\delta j^\mu$ of the current density $j^\mu=\langle\bar u\gamma^\mu u+\bar d\gamma^\mu d+\bar s\gamma^\mu s\rangle$.  Since we have taken the masses of $u$ and $d$ quarks to be the same and also assumed that the quark matter is isospin symmetric, we have $\sigma_u=\sigma_d\equiv\sigma_q$ and
obtain from Eq. (\ref{linres}) the following expressions:
\begin{widetext}
\begin{eqnarray}
\label{dll}
\delta\sigma_q(x)&=&-i\int d^4 \bar x \bigg\{\chi_{\sigma\sigma}(x-\bar x;M_q)[(-2G_S-2K_S\sigma_s(\bar x))\delta\sigma_q(\bar x)-2K_S\sigma_q(\bar x)\delta\sigma_s(\bar x)]+2G_V\chi^\mu_{\sigma j}(x-\bar x;M_q)\delta j_\mu(\bar x)\bigg\},\nonumber\\
\label{dss}
\delta\sigma_s(x)&=&-i\int d^4 \bar x \bigg\{\chi_{\sigma\sigma}(x-\bar x;M_s)\left[-4K_S\sigma_q(\bar x)\delta\sigma_q(\bar x)-2G_S\delta\sigma_s(\bar x)\right]+2G_V\chi^\mu_{\sigma j}(x-\bar x;M_s)\delta j_\mu(\bar x)\bigg\},\nonumber\\
\label{dj}
\delta j^\mu(x)&=&-i\int d^4 \bar x \bigg\{[2\chi^\mu_{\sigma j}(x-\bar x;M_q)(-2G_S-2K_S\sigma_s(\bar x))-4 K_S\chi^\mu_{\sigma j}(x-\bar x;M_s)\sigma_q(\bar x)]\delta\sigma_q(\bar x)-\left[2 G_S\chi^\mu_{\sigma j}(x-\bar x;M_s)\right.\nonumber\\
&&~~~~~~\left.-4K_S\chi^\mu_{\sigma j}(x-\bar x;M_q)\sigma_q(\bar x)\right]\delta\sigma_s(\bar x)+2G_V(2\chi_{jj}^{\mu\nu}(x-\bar x;M_q)+\chi_{jj}^{\mu\nu}(x-\bar x;M_s))\delta j_\nu(\bar x)\bigg\},
\end{eqnarray}
\end{widetext}
where
\begin{eqnarray}
\label{chi}
\chi_{\sigma\sigma}(x)&\equiv&\theta(t)\langle[\bar q(x)q(x),\bar q(0)q(0)]\rangle_0,\nonumber\\
\chi^\mu_{\sigma j}(x)&\equiv&\theta(t)\langle[\bar q(x)\gamma^\mu q(x),\bar q(0)q(0)]\rangle_0,\nonumber\\
\chi^{\mu\nu}_{jj}(x)&\equiv&\theta(t)\langle[\bar q(x)\gamma^\mu q(x),\bar q(0)\gamma^\nu q(0)]\rangle_0,
\end{eqnarray}
are quark correlators.

Taking the Fourier transformation of the equations in Eq. (\ref{dj}), we have
\begin{widetext}
\begin{eqnarray}
\label{dllk}
0&=&\left(1-i \widetilde\chi_{\sigma\sigma}(k;M_q)(2G_S+2K_S\sigma_s)\right)\delta\widetilde\sigma_q(k)-2K_Si \widetilde\chi_{\sigma\sigma}(k;M_q)\sigma_q\delta\widetilde\sigma_s(k)+2G_Vi\widetilde\chi^\mu_{\sigma j}(k;M_q)\delta \widetilde j_\mu(k),\nonumber\\
\label{dssk}
0&=&-4K_Si\widetilde\chi_{\sigma\sigma}(k;M_s)\sigma_q\delta\widetilde\sigma_q(k)+\left(1-2G_Si\widetilde\chi_{\sigma\sigma}(k;M_s)\right)\delta\widetilde\sigma_s(k)+2G_Vi\widetilde\chi^\mu_{\sigma j}(k;M_s)\delta \widetilde j_\mu(k),\nonumber\\
\label{djk}
0&=&-i\left(2\widetilde\chi^\mu_{\sigma j}(k;M_q)(2G_S+2K_S\sigma_s)+4 K_S\widetilde\chi^\mu_{\sigma j}(k;M_s)\sigma_q\right)\delta\widetilde\sigma_q(k)-i\left(4K_S\widetilde\chi^\mu_{\sigma j}(k;M_q)\sigma_q+2 G_S\widetilde\chi^\mu_{\sigma j}(k;M_s)\right)\delta\widetilde\sigma_s(k)\nonumber\\
&&+\left(g^{\mu\nu}+i 2G_V(2\widetilde\chi_{jj}^{\mu\nu}(k;M_q)+\widetilde\chi_{jj}^{\mu\nu}(k;M_s))\right)\delta\widetilde j_\nu(k),
\end{eqnarray}
\end{widetext}
with
\begin{equation}
\widetilde\chi(k)=\int d^4 x \chi(x)e^{ikx}.
\end{equation}

The equations in Eq.(\ref{djk}) have non-zero solutions if and only if
\begin{eqnarray}\label{dsp}
\mathrm{det}\left |
\begin{array}{ccc}
 A_{11} &  A_{12} &  A_{13}\\
 A_{21} &  A_{22} &  A_{23} \\
 A_{31} & A_{32} &  A_{33}
\end{array}
\right |=0,
\end{eqnarray}
where
\begin{eqnarray}
A_{11}&=&1- i\widetilde\chi_{\sigma\sigma}(k^0,\mathbf k;M_q)(2G_S+2K_S\sigma_s),\nonumber\\
A_{12}&=&-2iK_S \widetilde\chi_{\sigma\sigma}(k^0,\mathbf k;M_q)\sigma_q,\nonumber\\
A_{13}&=&2iG_V\widetilde\chi^\mu_{\sigma j}(k^0,\mathbf k;M_q),\nonumber\\
A_{21}&=&-4iK_S\widetilde\chi_{\sigma\sigma}(k^0,\mathbf k;M_s)\sigma_q,\nonumber\\
A_{22}&=&1-2iG_S\widetilde\chi_{\sigma\sigma}(k^0,\mathbf k;M_s),\nonumber\\
A_{23}&=&2iG_V\widetilde\chi^\mu_{\sigma j}(k^0,\mathbf k;M_s),\nonumber\\
A_{31}&=&-2i\widetilde\chi^\mu_{\sigma j}(k^0,\mathbf k;M_q)(2G_S+2K_S\sigma_s)\nonumber\\&&+4i K_S\widetilde\chi^\mu_{\sigma j}(k^0,\mathbf k;M_s)\sigma_q,\nonumber\\
A_{32}&=&-4iK_S\widetilde\chi^\mu_{\sigma j}(k^0,\mathbf k;M_q)\sigma_q+2i G_S\widetilde\chi^\mu_{\sigma j}(k^0,\mathbf k;M_s),\nonumber\\
A_{33}&=&g^{\mu\nu}+ i2G_V(2\widetilde\chi_{jj}^{\mu\nu}(k^0,\mathbf k;M_q)+\widetilde\chi_{jj}^{\mu\nu}(k^0,\mathbf k;M_s)).\nonumber\\
\end{eqnarray}

\subsection{Quark correlators}

In this subsection, we calculate the quark correlators defined in Eq. (\ref{chi}) in the imaginary time formalism. The retarded correlators $\widetilde\chi_{\sigma\sigma}$, $\widetilde\chi_{jj}$, and $\widetilde\chi_{\sigma j}$ are related to  the Matsubara correlator $\Pi$ by the Kubo-Martin-Schwinger (KMS) condition~\cite{Martin:1959jp} $\widetilde\chi(k)=-i\Pi(k^0+i0^+, \mathbf k)$, where
\begin{widetext}
\begin{eqnarray}
\label{Piss}
 \Pi_{\sigma\sigma}(i\nu_n,\mathbf k)&=&-T \sum_{a=1}^3\sum_{\omega_n}\int_{|\mathbf p|<\Lambda}^{|\mathbf {k+p}|<\Lambda}\frac{\mathrm d^3\mathbf p}{(2\pi)^3}\mathrm{Tr}[\widetilde S_a(i\omega_n,\mathbf p)\widetilde S_a(i\omega_n+i\nu_n,\mathbf k+\mathbf p)], \nonumber\\
\label{Pijj}
 \Pi^{\mu\nu}_{jj}(i\nu_n,\mathbf k)&=&-T\sum_{a=1}^3\sum_{\omega_n}\int_{|\mathbf p|<\Lambda}^{|\mathbf {k+p}|<\Lambda}\frac{\mathrm d^3\mathbf p}{(2\pi)^3}\mathrm{Tr}[\gamma^\mu \widetilde S_a(i\omega_n,\mathbf p)\gamma^\nu \widetilde S_a(i\omega_n+i\nu_n,\mathbf k+\mathbf p)],\nonumber\\
\label{Pisj}
 \Pi^{\mu}_{\sigma j}(i\nu_n,\mathbf k)&=&-T \sum_{a=1}^3\sum_{\omega_n}\int_{|\mathbf p|<\Lambda}^{|\mathbf {k+p}|<\Lambda}\frac{\mathrm d^3\mathbf p}{(2\pi)^3}\mathrm{Tr}[\widetilde S_a(i\omega_n,\mathbf p)\gamma^\mu \widetilde S_a(i\omega_n+i\nu_n,\mathbf k+\mathbf p)],
\end{eqnarray}
\end{widetext}
with the Matsubara frequency $\omega_n=(2n+1)\pi T$, $\nu_n=2n\pi T$, and $\widetilde S_a$ is the quark propagator in the imaginary time formalism. The subscript $a$ is the color index, and each color needs to be treated separately as the background color field in the PNJL model contributes differently to the chemical potentials of quarks of different colors.  Eq. (\ref{Pisj}) can be written in a more compact form as
\begin{eqnarray}
\Pi(i\nu_n,\mathbf k)&=&T \sum_{a=1}^3\sum_{\omega_n}\int_{|\mathbf p|<\Lambda}^{|\mathbf {k+p}|<\Lambda}\frac{\mathrm d^3\mathbf p}{(2\pi)^3}\mathrm{Tr}[\mathcal O_1\widetilde S_a(i\omega_n,\mathbf p)\nonumber\\&&
\mathcal O_2\widetilde S_a(i\omega_n+i\nu_n,\mathbf k+\mathbf p)],
\end{eqnarray}
where $\mathcal O_{1,2}$ denote the unit and the gamma matrices.

According to the KMS condition, the quark propagator $ \widetilde S_a(i\omega_n,\mathbf p)$  can be written in terms of the quark spectral function $\widetilde{\mathcal A}_a(p)$ as
\begin{equation}
\label{fermion}
\widetilde S_a(i\omega_n,\mathbf p)=-\int\frac{dp^0}{2\pi}\frac{\widetilde{\mathcal A}_a(p)}{i\omega_n-p^0},
\end{equation}
with
\begin{equation}
\widetilde{\mathcal A}_a(p)\equiv\int d^4x\mathcal A_a(x)e^{ipx/\hbar}\equiv\int d^4x \{q_a(x),\bar q_a(0)\}e^{ipx/\hbar}.
\end{equation}
Under the quasi-particle approximation, the quark spectral function can be written as
\begin{equation}
\widetilde{\mathcal A}_a(p)=\pi [\Delta_+(\mathbf p)\delta(p^0+\widetilde\mu_a-E_{\mathbf p})-\Delta_-(\mathbf p)\delta(p^0+\widetilde\mu_a+E_{\mathbf p})],
\end{equation}
where
\begin{equation}
\Delta_\pm(\mathbf p)=\pm \gamma^0-\frac{\mathbf p}{E_\mathbf{p}}\cdot\boldsymbol\gamma+\frac M {E_\mathbf{p}},
\end{equation}
and $\tilde\mu_{a}=\mu-2G_Vj^0+ i\phi_{aa}$ with $a=1,2,3$ are the effective chemical potentials. We therefore have
\begin{eqnarray}
\label{Pi}
&&\Pi(i\nu_n,\mathbf k)=\sum_{a=1}^3\int_{|\mathbf p|<\Lambda}^{|\mathbf {k+p}|<\Lambda}\frac{\mathrm d^3\mathbf p}{(2\pi)^3}\int\frac{dp^0}{2\pi}\frac{dp^{0\prime}}{2\pi}\nonumber\\
&&\quad\times T \sum_{\omega_n}\frac {-\mathrm{Tr}[\mathcal O_1\widetilde{\mathcal A}_a(p^0,\mathbf p)\mathcal O_2\widetilde{\mathcal A}_a(p^{0 \prime},\mathbf k+\mathbf p)]} {(i\omega_n-p^0)(i\omega_n+i\nu_n-p^{0\prime})}.
\end{eqnarray}
Evaluating the summation over the Matsubara frequency $\omega_n$ gives
\begin{widetext}
\begin{eqnarray}
\label{sumo}
&&T \sum_{\omega_n}\frac {-1} {(i\omega_n-p^0)(i\omega_n+i\nu_n-p^{0\prime})}=\frac {-1} {p^{0\prime}-p^0-i\nu_n}\left(T \sum_{\omega_n}\frac 1 {i\omega_n+i\nu_n-p^{0\prime}}-T \sum_{\omega_n}\frac 1 {i\omega_n-p^0}\right)\nonumber\\
&=&\frac {-1} {p^{0\prime}-p^0-i\nu_n}\frac 1 {2\pi i}\int_{-i\infty+0^+}^{i\infty+0^+} dz
\bigg(\frac 1 {z+i\nu_n-p^{0\prime}}+\frac 1 {-z+i\nu_n-p^{0\prime}}-\frac 1 {z-p^0}-\frac 1 {-z-p^0}\bigg)\times\left(\frac 1 2-f_0(z)\right)\nonumber\\
&=& \frac {f_0(p^0)-f_0(p^{0\prime})} {p^{0\prime}-p^0-i\nu_n},
\end{eqnarray}
\end{widetext}
where $f_0(x)=(\exp(x/T)+1)^{-1}$ is the Fermi-Dirac distribution function. Notice that the function $1/2-f_0(z)$ has simple poles at $z=(2n+1)\pi i T$ with same residue $T$. The second equality in Eq. (\ref{sumo}) is obtained after counting these residues, while the third equality follows after using the relation $f_0(x+i\nu_n)=f_0(x)$. 

We have so far taken the Planck constant $\hbar=1$. Including explicitly $\hbar$ in Eq. (\ref{Pi}) gives
\begin{widetext}
\begin{eqnarray}
\label{chitilde}
&&i\widetilde\chi(\omega,\mathbf k)=\Pi(\hbar\omega+i0^+, \mathbf k)\nonumber\\
 &=&\frac 1 4\sum_{a=1}^3\int_{|\mathbf p|<\Lambda}^{|\hbar\mathbf k+\mathbf p|<\Lambda}\frac{\mathrm d^3\mathbf p}{(2\pi)^3}\bigg\{
\mathrm{Tr}[\mathcal O_1\Delta_+(\mathbf p)\mathcal O_2\Delta_+(\hbar\mathbf k+\mathbf p)]
\frac {f_0(E_\mathbf{p}-\widetilde\mu_a)-f_0(E_{\mathbf p+\hbar \mathbf k}-\widetilde\mu_a)} {E_{\mathbf p+\hbar \mathbf k}-E_\mathbf{p}-\hbar\omega-i0^+}\nonumber\\
&&+\mathrm{Tr}[\mathcal O_1\Delta_-(\mathbf p)\mathcal O_2\Delta_-(\hbar\mathbf k+\mathbf p)]
\frac {f_0(E_{\mathbf p+\hbar \mathbf k}+\widetilde\mu_a)-f_0(E_\mathbf{p}+\widetilde\mu_a)} {E_\mathbf{p}-E_{\mathbf p+\hbar \mathbf k}-\hbar\omega-i0^+}\nonumber\\&&-\mathrm{Tr}[\mathcal O_1\Delta_+(\mathbf p)\mathcal O_2\Delta_-(\hbar\mathbf k+\mathbf p)]
\frac {f_0(E_{\mathbf p+\hbar \mathbf k}+\widetilde\mu_a)+f_0(E_\mathbf{p}-\widetilde\mu_a)-1} {-E_\mathbf{p}-E_{\mathbf p+\hbar \mathbf k}-\hbar\omega-i0^+}\nonumber\\
&&~~-\mathrm{Tr}[\mathcal O_1\Delta_-(\mathbf p)\mathcal O_2\Delta_+(\hbar\mathbf k+\mathbf p)]
\frac {1-f_0(E_{\mathbf p+\hbar \mathbf k}-\widetilde\mu_a)-f_0(E_\mathbf{p}+\widetilde\mu_a)} {E_\mathbf{p}+E_{\mathbf p+\hbar \mathbf k}-\hbar\omega-i0^+}\bigg\}.
\end{eqnarray}
\end{widetext}
After making the variable change $\mathbf p^\prime=-\mathbf p-\hbar\mathbf k$ in terms containing $f_0(E_{\mathbf p+\hbar\mathbf k}\pm\widetilde\mu^a)$, which also satisfies $|\mathbf p^\prime|<\Lambda$ and $|\mathbf p^\prime+\hbar\mathbf k|<\Lambda$, and writing ${\bf p}^\prime$ again as ${\bf p}$, the above equation can be rewritten as
\begin{widetext}
\begin{eqnarray}
\label{chi2}
i\widetilde\chi(\omega,\mathbf k)
 &=&\frac 1 4\sum_{a=1}^3\int_{|\mathbf p|<\Lambda}^{|\hbar\mathbf k+\mathbf p|<\Lambda}\frac{\mathrm d^3\mathbf p}{(2\pi)^3}\bigg\{
\bigg(\frac{\mathrm{Tr}[\mathcal O_1\Delta_+(\mathbf p)\mathcal O_2\Delta_+(\hbar\mathbf k+\mathbf p)]}{E_{\mathbf p+\hbar \mathbf k}-E_\mathbf{p}-\hbar\omega-i0^+}
+\frac{\mathrm{Tr}[\mathcal O_1\Delta_+(-\mathbf p-\hbar\mathbf k)\mathcal O_2\Delta_+(-\mathbf p)]}{E_{\mathbf p+\hbar \mathbf k}-E_\mathbf{p}+\hbar\omega+i0^+}\nonumber\\
&&+\frac{\mathrm{Tr}[\mathcal O_1\Delta_+(\mathbf p)\mathcal O_2\Delta_-(\hbar\mathbf k+\mathbf p)]}{E_\mathbf{p}+E_{\mathbf p+\hbar \mathbf k}+\hbar\omega+i0^+}
+\frac{\mathrm{Tr}[\mathcal O_1\Delta_-(-\mathbf p-\hbar\mathbf k)\mathcal O_2\Delta_+(-\mathbf p)]}{E_\mathbf{p}+E_{\mathbf p+\hbar \mathbf k}-\hbar\omega-i0^+}
\bigg)
f_0(E_\mathbf{p}-\widetilde\mu_a)\nonumber\\
&&+\bigg(\frac{\mathrm{Tr}[\mathcal O_1\Delta_-(-\mathbf p-\hbar\mathbf k)\mathcal O_2\Delta_-(-\mathbf p)]}{E_\mathbf{p+\hbar \mathbf k}-E_{\mathbf p}-\hbar\omega-i0^+}
+\frac{\mathrm{Tr}[\mathcal O_1\Delta_-(\mathbf p)\mathcal O_2\Delta_-(\hbar\mathbf k+\mathbf p)]}{E_{\mathbf p+\hbar \mathbf k}-E_\mathbf{p}+\hbar\omega+i0^+} \nonumber\\
&&+\frac{\mathrm{Tr}[\mathcal O_1\Delta_-(\mathbf p)\mathcal O_2\Delta_+(\hbar\mathbf k+\mathbf p)]}{E_\mathbf{p}+E_{\mathbf p+\hbar \mathbf k}-\hbar\omega+i0^+}
+\frac{\mathrm{Tr}[\mathcal O_1\Delta_+(-\mathbf p-\hbar\mathbf k)\mathcal O_2\Delta_-(-\mathbf p)]}{E_\mathbf{p}+E_{\mathbf p+\hbar \mathbf k}+\hbar\omega+i0^+}\bigg)
f_0(E_\mathbf{p}+\widetilde\mu_a) \nonumber\\
&&-\frac{\mathrm{Tr}[\mathcal O_1\Delta_-(\mathbf p)\mathcal O_2\Delta_+(\hbar\mathbf k+\mathbf p)]}{E_\mathbf{p}+E_{\mathbf p+\hbar \mathbf k}-\hbar\omega+i0^+}
-\frac{\mathrm{Tr}[\mathcal O_1\Delta_+(-\mathbf p-\hbar\mathbf k)\mathcal O_2\Delta_-(-\mathbf p)]}{E_\mathbf{p}+E_{\mathbf p+\hbar \mathbf k}+\hbar\omega+i0^+}\bigg\}.
\end{eqnarray}
\end{widetext}
We note that if $\mathcal O_{1,2}$ are such that $\mathrm{Tr}[\mathcal O_1\Delta_\pm(\mathbf p)\mathcal O_2\Delta_\pm(\hbar\mathbf k+\mathbf p)]= \mathrm{Tr}[\mathcal O_1\Delta_\pm(-\mathbf p-\hbar\mathbf k)\mathcal O_2\Delta_\pm(-\mathbf p)]$ and $\mathrm{Tr}[\mathcal O_1\Delta_\pm(\mathbf p)\mathcal O_2\Delta_\mp(\hbar\mathbf k+\mathbf p)]=\mathrm{Tr}[\mathcal O_1\Delta_\mp(-\mathbf p-\hbar\mathbf k)\mathcal O_2\Delta_\pm(-\mathbf p)]$, then $\widetilde\chi(\omega,\mathbf k)=\widetilde\chi(-\omega,\mathbf k)$. On the other hand, if $\mathrm{Tr}[\mathcal O_1\Delta_\pm(\mathbf p)\mathcal O_2\Delta_\pm(\hbar\mathbf k+\mathbf p)]= -\mathrm{Tr}[\mathcal O_1\Delta_\pm(-\mathbf p-\hbar\mathbf k)\mathcal O_2\Delta_\pm(-\mathbf p)]$ and $\mathrm{Tr}[\mathcal O_1\Delta_\pm(\mathbf p)\mathcal O_2\Delta_\mp(\hbar\mathbf k+\mathbf p)]=-\mathrm{Tr}[\mathcal O_1\Delta_\mp(-\mathbf p-\hbar\mathbf k)\mathcal O_2\Delta_\pm(-\mathbf p)]$, then $\widetilde\chi(\omega,\mathbf k)=-\widetilde\chi(-\omega,\mathbf k)$.
Furthermore, taking the fluctuation in the current density as a longitudinal wave, i.e., $\delta \mathbf j=\delta j_z \hat\mathbf k$, then terms such as $\widetilde\chi^{0x}_{jj}$ vanish. Among the remaining quark correlators, $\widetilde\chi_{\sigma\sigma}(\omega,\mathbf k)$, $\widetilde\chi^0_{\sigma j}(\omega,\mathbf k)$, $\widetilde\chi_{jj}^{00}(\omega,\mathbf k)$, and $\widetilde\chi_{jj}^{zz}(\omega,\mathbf k)$ are even in $\omega$, while $\widetilde\chi^z_{\sigma j}(\omega,\mathbf k)$ and $\widetilde\chi_{jj}^{0z}(\omega,\mathbf k)$ are odd in $\omega$. The determinant in Eq. (\ref{dsp}) is thus even in $\omega$, i.e., both $\omega=\omega_\mathbf k$ and $\omega=-\omega_\mathbf k$ are solutions. For unstable modes, corresponding to imaginary $\omega$, i.e., $\omega_\mathbf k=i\Gamma_k$, the imaginary frequency $\omega= i\Gamma_k$ and $\omega=-i\Gamma_k$ then correspond to unstable modes that grow and decay exponentially in time with a growth or decay rate $\Gamma_k$. 

\subsection{Results}

\begin{figure}[htb]
\vspace{-0cm}
\includegraphics[width=0.4 \textwidth]{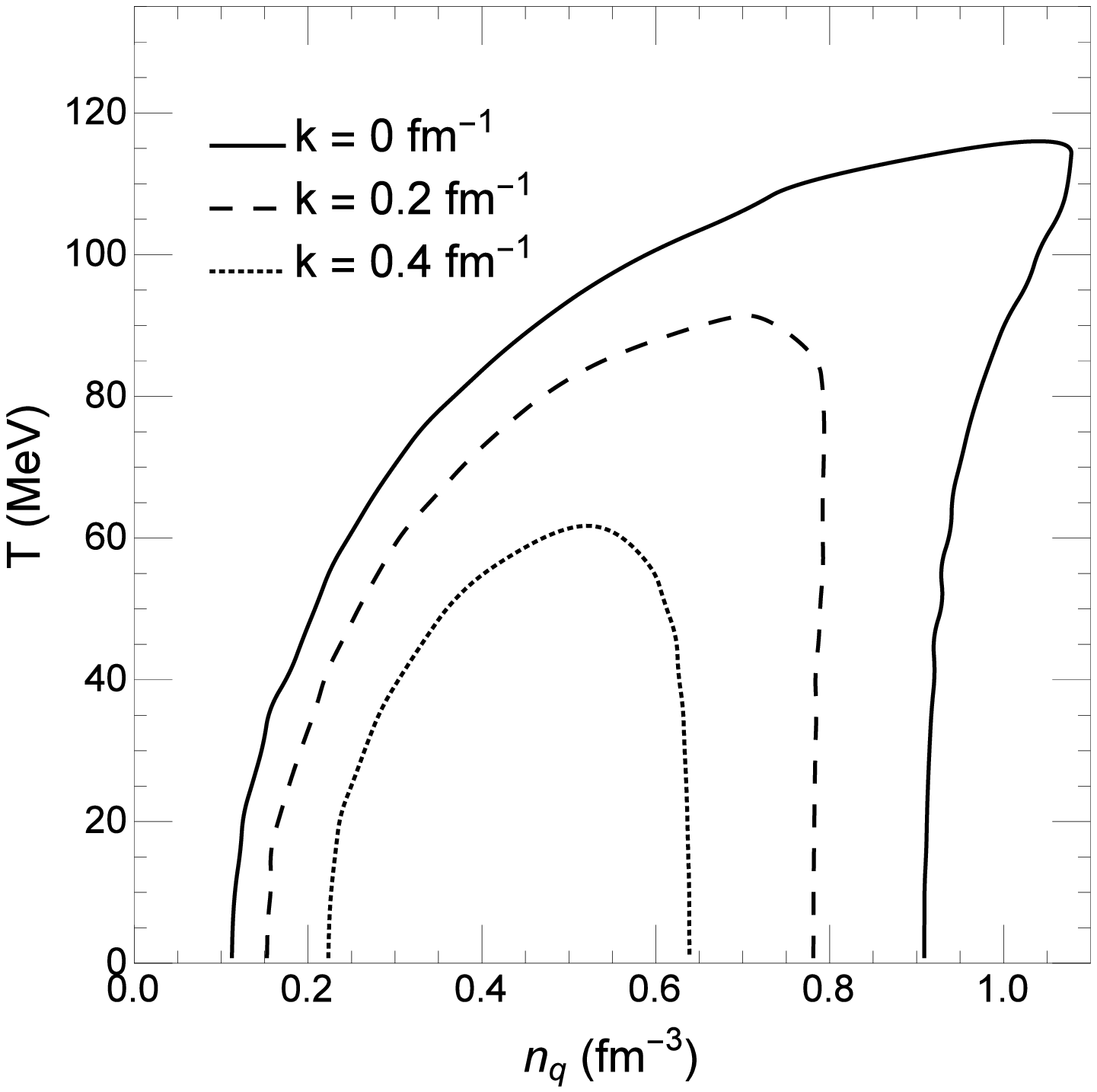}
\caption{Spinodal boundaries of unstable modes of different wave numbers in the temperature and net quark density plane from the PNJL model with $G_V=0$.}
\label{boundaries_Gv_00_PNJL}
\end{figure}

\begin{figure}[htb]
\vspace{-0cm}
\includegraphics[width=0.4 \textwidth]{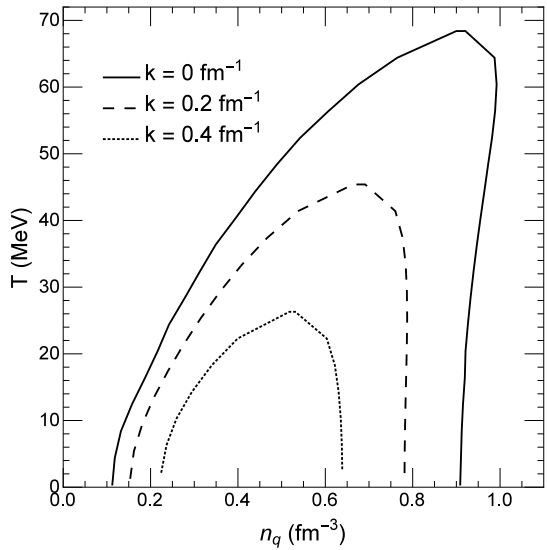}
\caption{Spinodal boundaries of unstable modes of different wave numbers in the temperature and net quark density plane from the NJL model with $G_V=0$.}
\label{boundaries_Gv_00_NJL}
\end{figure}

In this subsection, we show results obtained from Eq. (\ref{dsp}) for the spinodal boundaries of unstable modes of different wave numbers. Since the determinant in Eq. (\ref{dsp}) is even in $k^0$, it has a minimum at $k^0=0$. Therefore, for a given wave vector $\mathbf k$, temperature $T$, and net baryon density $n_q$, Eq. (\ref{dsp}) can be solved if and only if the determinant is negative or zero when $k_0=0$. We can thus obtain the boundaries of the spinodal instability region for different $\mathbf k$ by solving Eq. (\ref{dsp}) with $k^0=0$, and they are shown in Figs.~\ref{boundaries_Gv_00_PNJL} and \ref{boundaries_Gv_00_NJL} with $G_V=0$ but with (PNJL) and without (NJL) the Polyakov loop, respectively. It is seen that for unstable modes of same wave number, the spinodal instability region is larger in the PNJL than in the NJL model due to the effect of the Polyakov loop.  The highest temperature $T_c$ of the spinodal instability region  is about $68~\mathrm{MeV}$ in the NJL model and $120~\mathrm{MeV}$ in the PNJL model.  Comparing these results with those shown in Figs. \ref{vs0_PNJL} and \ref{vs0_NJL} based on the thermodynamic approach, we find that the spinodal boundaries of unstable modes of $k=0$ coincide with the boundaries determined from $(\partial_n p)_T\leq 0$ but are different from those determined from  $(\partial_n p)_S\leq 0$.  Therefore, unstable modes in the long wavelength limit correspond to the isothermal spinodal instability, and this is because the time evolution operator $\mathcal U(t,t_0)$ becomes non-unitary after linearization and the entropy $S=\mathrm{tr}(\rho\ln\rho)$ of the system thus no longer remains constant.
   
Figs. \ref{boundaries_Gv_00_PNJL} and \ref{boundaries_Gv_00_NJL} also show that the spinodal instability region shrinks as the wave number of unstable modes increases or its wavelength becomes shorter. This is because clumps of quark matter or the high density regions are more likely to merge into lager clumps, which correspond to modes of smaller wave number or longer wavelength, and this effect is larger at higher temperature.  We note that this effect becomes less important in the semiclassical case, where the chance for a particle to move from a small cluster to a larger cluster is the same as that of the reverse process, due to the zero range nature of the interactions in the PNJL and NJL models as shown in the next section for the case of the NJL model.

\begin{figure}[htb]
	\vspace{-0cm}
	\includegraphics[width=0.4 \textwidth]{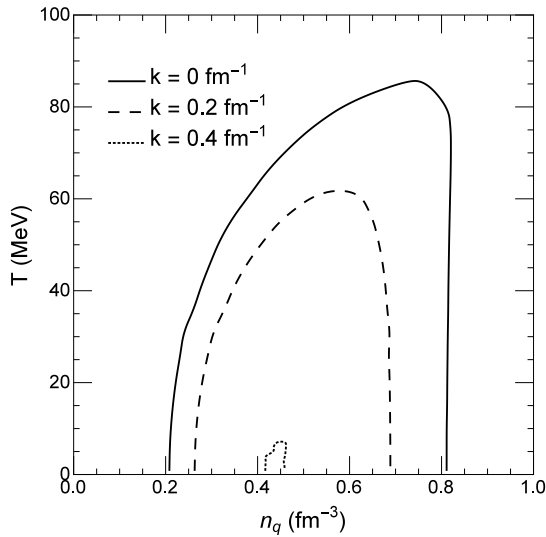}
	\caption{Spinodal boundaries of different wave numbers in the temperature and net quark density plane from the PNJL model with $G_V=0.2~G_S$.}
	\label{boundaries_Gv_02_PNJL}
\end{figure}

\begin{figure}[htb]
	\vspace{-0cm}
	\includegraphics[width=0.4 \textwidth]{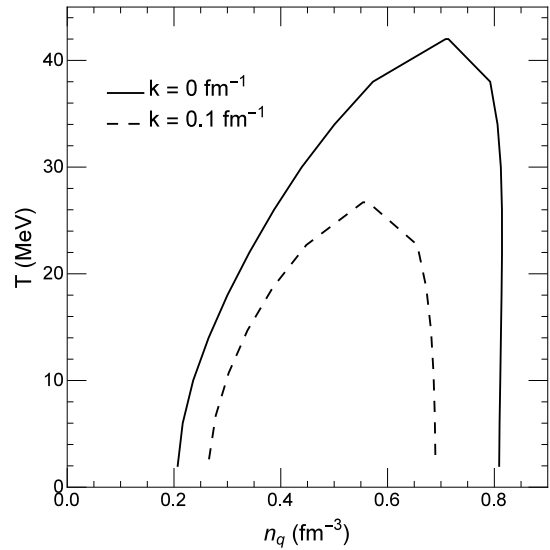}
	\caption{Spinodal boundaries of unstable modes of different wave numbers in the temperature and net quark density plane from the NJL model with $G_V=0.2~G_S$.}
	\label{boundaries_Gv_02_NJL}
\end{figure}

We have also studied the effect of the vector interaction on the spinodal boundaries of unstable modes of different wave numbers by using $G_V=0.2~G_S$ in the PNJL and NJL models, and the results are shown in Figs. \ref{boundaries_Gv_02_PNJL} and \ref{boundaries_Gv_02_NJL}, respectively. It is seen that the vector interaction shrinks the unstable region, particularly for unstable modes of large wave number or shorter wavelength.  For example, Fig.~\ref{boundaries_Gv_02_NJL} shows that unstable modes with $k\geq 0.2~\mathrm{fm}^{-1}$  disappear for $G_V=0.2~G_S$. This agrees with the expectation that the repulsive interaction drives particles away from clumps and thus destroys unstable modes.

\begin{figure}[htb]
\vspace{-0cm}
\includegraphics[width=0.5 \textwidth]{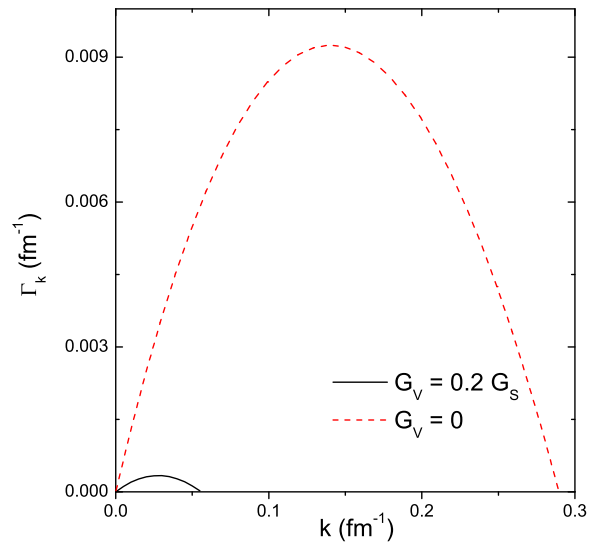}
\caption{(Color online) Dispersion relation of unstable modes in a quark matter of net quark density $n_q=0.7$ fm$^{-3}$ and temperature $T=70$ MeV for both $G_V=0$  and  $G_V=0.2~G_S$ based on the PNJL model. }
\label{dispersion}
\end{figure}

We further show in Fig.~\ref{dispersion} the dispersion relation of unustable modes, i.e.,  its growth rate $\Gamma_k$ as a function of the wave number $k$, in a quark matter of net quark density $n_q=0.7~\mathrm{fm}^{-3}$ and at temperature $T=70$ MeV, for the case of $G_V=0$ (dashed line) and the case of $G_V=0.2~G_S$ (solid line).  The vector interaction is seen to dramatically reduce the growth rate of unstable modes.  For $G_V=0$, the growth rate peaks at $k=0.15~\mathrm{fm^{-1}}$, implying that the size of the most likely quark clumps due to density fluctuations is about $2\pi/k_{\rm max}\sim 40~ \mathrm{fm}$. The typical growth rate is $0.01 \mathrm{fm}^{-1}$, indicating that it takes about 100 fm for the fluctuations to grow. This time duration is an order of magnitude longer than the typical lifetime of a heavy-ion collision, which is about 10 fm, making the effect of instabilities hardly visible.  The instabilities can be enhanced by increasing the attraction interaction between quarks, namely increasing the values of $G_S$ and $K_S$. The latter then requires a larger cutoff parameter $\Lambda$ in order to reproduce the correct meson masses in vacuum. Also, nonlinear effects, which are neglected in the linear response theory, may enhance the growth rate of unstable modes and result in appreciable density fluctuations in a much shorter time.

The above growth rate of unstable modes is smaller than that obtained in Ref.~\cite{PhysRevC.87.054903} based on the quark-meson coupling model.  This is due to a smaller pressure difference in our model. To see this, we note that the growth rate is proportional to the speed of sound of a medium, given by $v_S^2=\frac{\partial p}{\partial n}\frac n{p+\epsilon}$. In the PNJL model, the pressure at $T=0$ decreases by 20 MeV/fm$^{-3}$ when the density increases by 0.8  fm $^{-3}$. In Ref.~\cite{PhysRevC.87.054903}, the pressure decreases by about 200 MeV/fm$^{3}$ as the density increases by about 0.3 fm$^{-3}$, thus resulting in a factor of 4 reduction of the growth rate in our model.  Also, quantum effects, which suppress unstable modes of shorter wavelengths and thus  reduce the growth rate, are not taken into account in Ref.~\cite{PhysRevC.87.054903} and are also responsible for the smaller growth rate obtained in our study.
 
\section{The linearized Boltzmann equation: collisional effects}

Higher-order corrections to the correlators in Eq.(\ref{Piss}) obtained in the linear response theory can in principle be included but is much more involved. For simplicity, we study their effects on the growth rate of unstable modes by using the linearized transport (or Boltzmann) equation, which agrees with the linear response approach in the semiclassical limit.  Another advantage of this approach is that the transport equation can be solved by the test particle method and thus easily used to study unstable modes of large amplitude, which we leave as a future study. In the present study, we consider the case of NJL model with vanishing vector interaction, as we already know from the previous section that the vector interaction suppresses spinodal instabilities. The reason for not studying the collisional effect based on the PNJL model is because a consistent treatment of collisional terms in the transport equation is currently not known. 

\subsection{The theoretic framework}

The Boltzmann equation can be written in a concise form~\cite{Song:2012cd},
\begin{equation}
D[f_a]=C[f_z],
\end{equation}
in terms of the drift term
\begin{equation}
D[f_a]\equiv \partial_t f_a+\mathbf v\cdot \nabla_\mathbf{r} f_a+\frac{M_a}{E_a}\nabla_\mathbf{r} V^S\nabla_\mathbf p f_a,
\end{equation}
where $\mathbf v=\mathbf p/E_\mathbf p$ is the velocity, and the collision term
\begin{widetext}
\begin{eqnarray}
C[f_a]&\equiv& \sum_{bcd}\frac 1 {1+\delta_{ab}}\int\frac{d^3\mathbf p_b}{(2\pi)^32E_b}\frac{d^3\mathbf p_c}{(2\pi)^32E_c}\frac{d^3\mathbf p_d}{(2\pi)^32E_d}\frac{(2\pi)^4}{2E_a}\delta^4(p_a+p_b-p_c-p_d)\nonumber\\
&&\times|\mathcal M_{ab}|^2\left[f_cf_d(1-f_a)(1-f_b)-f_af_b(1-f_c)(1-f_d)\right]
\end{eqnarray}
\end{widetext}
that describes scatterings among quarks. In the above, the subscriptions $a$, $b$, $c$, and $d$ denote the spin, flavor, color, and baryon charge (quark or anti-quark) of a parton.

Expanding the distribution function $f$ around its equilibrium value $f^0$, which satisfies the condition $D[f^0_a]=C[f^0_a]=0$, by writing $f=f^0+\delta f$ and keeping only terms linear in $\delta f$, we obtain
\begin{equation}
\label{D}
D[f_a]\approx \partial_t \delta f_a+\mathbf v\cdot \nabla_\mathbf{r} \delta f_a+\frac{M_a}{E_a}\nabla_\mathbf{r} \delta V^S\nabla_\mathbf p f^0_a,
\end{equation}
where we have introduced $V^S=V_0^S+\delta V^S$ and used the fact that $\nabla_\mathrm{r} V^S_0=0$ at equilibrium.  Similarly, the collision term becomes
\begin{widetext}
\begin{eqnarray}
\label{c1}
&&C[f_a]\approx \sum_{bcd}\int d^3\mathbf p_b d^3\mathbf p_c d^3\mathbf p_d\frac{(2\pi)^{-5}\delta^4(p_a+p_b-p_c-p_d)}{(1+\delta_{ab})2E_a2E_b2E_c2E_d}|\mathcal M_{ab}|^2 f_a^0 f_b^0 (1-f_c^0) \nonumber\\
 &&~~~~~~ (1-f_d^0)\left[-\frac{\delta f_a}{f_a^0(1-f_a^0)}-\frac{\delta f_b}{f_b^0(1-f_b^0)}+\frac{\delta f_c}{f_c^0(1-f_c^0)}+\frac{\delta f_d}{f_d^0(1-f_d^0)}\right].
\end{eqnarray}
\end{widetext}
Neglecting the contributions from $\delta f_b$, $\delta f_c$, and $\delta f_d$ results in the so-called relaxation approximation to the collisional term:
\begin{equation}
\label{C}
C[f_a]\approx -\frac 1 {\tau_a}\delta f_a,
\end{equation}
where the relaxation time $\tau_a$ characterizes the time for the system to evolve from a non-equilibrium state to an equilibrium one and is given by
\begin{widetext}
\begin{eqnarray}
\label{tau1}
\frac 1 {\tau_a} = \sum_{bcd}\int d^3\mathbf p_b d^3\mathbf p_c d^3\mathbf p_d\frac{(2\pi)^{-5}\delta^4(p_a+p_b-p_c-p_d)}{(1+\delta_{ab})2E_a2E_b2E_c2E_d}|\mathcal M_{ab}|^2 \frac{f_b^0 (1-f_c^0)(1-f_d^0)}{1-f_a^0}.\nonumber\\
\end{eqnarray}
\end{widetext}
To evaluate the integrals in the above equation, we introduce the following change of variables: $\mathbf P=\mathbf p_a+\mathbf p_b$, $\mathbf P^\prime=\mathbf p_c+\mathbf p_d$, and $\mathbf p=\mathbf p_c-\mathbf p_d$. Using the relation $d^3\mathbf p_b d^3\mathbf p_c d^3\mathbf p_d=2^{-3}d^3\mathbf P d^3\mathbf P^\prime d^3 \mathbf p$ and ${\bf P}={\bf P}^\prime$, Eq.(\ref{tau1}) becomes
\begin{widetext}
\begin{equation}
\frac 1 {\tau_a}=\sum_{bcd}\int d^3\mathbf P d^3\mathbf p \frac{(2\pi)^{-5}\delta(E_a+E_b-E_c-E_d)}{(1+\delta_{ab})2^3 2E_a2E_b2E_c2E_d}|\mathcal M_{ab}|^2 \frac{f_b^0 (1-f_c^0)(1-f_d^0)}{1-f_a^0}.
\end{equation}
\end{widetext}
Writing
\begin{equation}
\delta(E_a+E_b-E_c-E_d)=\int dE \delta(E_a+E_b-E)\delta(E_c+E_d-E),
\end{equation}
it is then easy to show that for the first $\delta$ function, we have
\begin{equation}
\delta(E-E_a-E_b)=\frac{E_b}{P p_a}\delta\left(x-\frac{p_a^2-p_b^2+P^2}{2Pp_a}\right),
\end{equation}
where $x$ denotes $\cos\angle(\mathbf P,\mathbf p_a)$. The second $\delta$ function is tedious to evaluate unless the colliding particles all have same mass. Since we are interested in quark matter that has temperatures below the critical temperature $T_c$ and net baryon chemical potentials smaller than $1$ GeV, very few strange (anti-)quarks are present and the scatterings are mostly among light $u$ and $d$ quarks of similar masses. Taking the equal mass limit, the second $\delta$ function can be expressed as
\begin{equation}
\delta(E_c+E_d-E)=\delta\left(p-E\sqrt{\frac{s-2m^2}{E^2-P^2x^{\prime2}}}\right)\frac{4EE_cE_d}{p(E^2-P^2x^{\prime2})},
\end{equation}
with $x=\cos\angle(\mathbf P,\mathbf p)$ and $s=(p_a+p_b)^2$. Assuming that the scattering cross sections are isotropic, then $|\mathcal M_{ab}|^2=16\pi s\sigma^{ab}_\mathrm{CM}$, and this leads to
\begin{widetext}
\begin{eqnarray}
\frac {1}{\tau_a}=\sum_{bcd}\frac 1 {(1+\delta_{ab})2(2\pi)^2p_aE_a(1-f^0_a)}\int dE dP dx^\prime s\sigma^{ab}_\mathrm{CM} \frac{PE^2}{E^2-P^2x^{\prime2}}\sqrt{\frac{s-2m^2}{E^2-P^2x^{\prime2}}}f_b^0 (1-f_c^0)(1-f_d^0).
\end{eqnarray}
\end{widetext}
For collisions of $u$ and $d$ quarks only, the summation in the above equation becomes a constant factor: $\sum_{bcd}(1+\delta_{ab})^{-1}=2~(\mathrm{spins})\times3~(\mathrm{colors})\times2~(\mathrm{flavors)}-1/2~(\mathrm{identical~particle})=11.5$~\cite{Chakraborty:2010fr}. 

In the relaxation approximation to the collision term, the linearized transport equation can now be expressed as
\begin{equation}
\label{LBZM1}
\partial_t \delta f_a+\mathbf v\cdot \nabla_\mathbf{r} \delta f_a++\frac{M_a}{E_a}\nabla_\mathbf{r} \delta V^S_a\nabla_\mathbf p f^0_a+\tau^{-1}_a\delta f_a=0.
\end{equation}
In terms of the Fourier transform of $\delta f$ and $\delta V^S$,
\begin{eqnarray}
\delta\tilde f(\mathbf k, \mathbf p, \omega)&=&\int dt d^3\mathbf x \delta f(\mathbf x, \mathbf p, t)\exp(i\omega t-i\mathbf k\cdot\mathbf x),\nonumber\\
\delta\tilde V^S(\mathbf k, \omega)&=&\int dt d^3\mathbf x \delta V^S(\mathbf x, t)\exp(i\omega t-i\mathbf k\cdot\mathbf x).\nonumber\\
\end{eqnarray}
Eq. (\ref{LBZM1}) can be rewritten as
\begin{equation}
\label{LBZM2}
(\omega+i\tau^{-1}_a-\mathbf k\cdot\mathbf v_a) \delta\tilde f_a+\frac{M_a}{E_a}\delta\tilde V^S_a\mathbf k\cdot\nabla_\mathbf p f^0_a=0.
\end{equation}
Since $V^S_i=2G_S\langle \bar qq\rangle_i+2K\langle\bar qq\rangle_j\langle\bar qq\rangle_k$ with $i,j,k$ indicating the flavors and $i\neq j\neq k$ as shown in the previous section, their variations are thus given by
\begin{eqnarray}
\label{dMq}
\delta V^S_q&=&(2G_S+2K\langle\bar ss\rangle)\delta\langle\bar qq\rangle+2K\langle\bar qq\rangle\delta\langle\bar ss\rangle,
\nonumber\\
\label{dMs}
\delta V^S_s&=&4K\langle\bar qq\rangle\delta\langle\bar qq\rangle+2G_S\delta\langle\bar ss\rangle,
\end{eqnarray}
with $q$ denoting $u$ or $d$ quark. For an isospin symmetric quark matter considered in the present study, $u$ and $d$ quarks have same mass and condensate, and we then  obtain from Eq. (\ref{LBZM2}) the following:
\begin{widetext}
\begin{eqnarray}
\label{dfq}
\delta f_{q,\bar q}&=&\frac{2\mathbf k\cdot\nabla_\mathbf{p} f^0_{q,\bar q} M_q/E_q[(G_S+K\langle\bar ss\rangle)\delta\langle\bar qq\rangle+K\langle\bar qq\rangle\delta\langle\bar ss\rangle]}{\omega+i\tau_q^{-1}-\mathbf k\cdot\mathbf v},\nonumber\\
\label{dfs}
\delta f_{s,\bar s}&=&\frac{2\mathbf k\cdot\nabla_\mathbf{p} f^0_{q,\bar q} M_q/E_q[2K\langle\bar qq\rangle\delta\langle\bar qq\rangle+G_S\delta\langle\bar ss\rangle]}{\omega-\mathbf k\cdot\mathbf v}.
\end{eqnarray}
\end{widetext}
Expressing both $\delta\langle\bar qq\rangle$ and $\delta\langle\bar ss\rangle$ in terms of $\delta f_{q,\bar q}$ and $\delta f_{s,\bar s}$ according to
\begin{widetext}
\begin{eqnarray}
\label{dqq}
\delta\langle\bar qq\rangle&=&2N_c\int\frac{d^3\mathbf p}{(2\pi)^3}\left(\frac{M_q}{E_q}(\delta f_q+\delta f_{\bar q})+\frac{p^2}{E_q^3}(f^0_q+f^0_{\bar q}-1)\delta M_q\right),\nonumber\\
\label{dss}
\delta\langle\bar ss\rangle&=&2N_c\int\frac{d^3\mathbf p}{(2\pi)^3}\left(\frac{M_s}{E_s}(\delta f_s+\delta f_{\bar s})+\frac{p^2}{E_s^3}(f^0_s+f^0_{\bar s}-1)\delta M_s\right),
\end{eqnarray}
\end{widetext}
and substituting $\delta\tilde f_{q,\bar q}$, $\delta\tilde f_{s,\bar s}$, $\delta M_q$, and $\delta M_s$ in Eq. (\ref{dqq}), we obtain after some simplifications the following results:
\begin{widetext}
\begin{equation}
\label{X1}
\left(\begin{array}{cc}
1-2(G_S+K\langle\bar ss\rangle)(\chi_q-\xi_q) & -2K\langle\bar qq\rangle(\chi_q-\xi_q)\\
-4K\langle\bar qq\rangle(\chi_s-\xi_s) & 1-2G_S(\chi_s-\xi_s)
\end{array}\right)
\left(\begin{array}{c}
\delta\langle\bar qq\rangle\\
\delta\langle\bar ss\rangle
\end{array}\right)=0,
\end{equation}
\end{widetext}
where
\begin{widetext}
\begin{eqnarray}
\label{chi3}
\chi&=&2N_c\int\frac{d^3\mathbf p}{(2\pi)^3}\left(\frac M E\right)^2\frac{\mathbf k\cdot\nabla_\mathbf{p}(f^0+\bar f^0)}{\omega+i\tau^{-1}-\mathbf k\cdot\mathbf v}\nonumber\\
&=&\frac{N_cM^2}{\pi^2T}\int dE v\left[f^0(f^0-1)+\bar f^0(\bar f^0-1)\right]\left(\frac{\tau^{-1}-i\omega}{kv}\arctan\frac{kv}{\tau^{-1}-i\omega}-1\right),\nonumber\\
\end{eqnarray}
\end{widetext}
and
\begin{equation}
\label{chi4}
\xi=\frac{N_c}{\pi^2}\int dE\frac{p^3}{E^2}(f^0+\bar f^0-1).
\end{equation}
We note that Eq.(\ref{chitilde}) reduces to Eq.(\ref{chi3}) in the limit $\hbar=0$. Since $\delta\langle\bar qq\rangle$ and $\delta\langle\bar ss\rangle$ can be of any value, Eq. (\ref{X1}) is satisfied if and only if:
\begin{equation}
\label{X2}
\left |\begin{array}{cc}
1-2(G_S+K\langle\bar ss\rangle)(\chi_q-\xi_q) & -2K\langle\bar qq\rangle(\chi_q-\xi_q)\\
-4K\langle\bar qq\rangle(\chi_s-\xi_s) & 1-2G_S(\chi_s-\xi_s)
\end{array}\right |=0.
\end{equation}
By solving eq. (\ref{X2}), we can obtain the relation between the frequency and wave number of collective modes in quark matter, i.e., its dispersion relation $\omega(k)$. These collective modes become unstable and grow with time if their frequencies are imaginary, i.e., $\omega=i\Gamma_k$, which can occur in quark matter for some temperatures and densities as discussed in the previous chapter. Since $\chi(\omega,k) =\chi(\omega/k)$ when $\tau^{-1}=0$, the growth rate $\Gamma_k$ is thus proportional to $k$ in the absence of collisions. This is in contrast to the results obtained in the quantum linear response theory of Section III where the growth rate of unstable modes of larger wave number is suppressed.   

\subsection{Results}

\begin{figure}[htbp]
\vspace{-0cm}
\includegraphics[width=0.4 \textwidth]{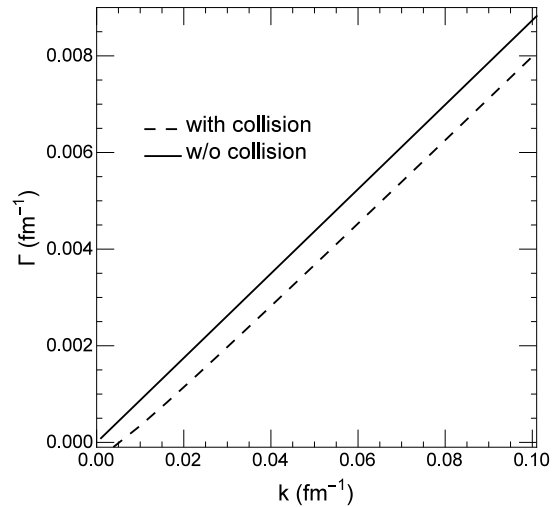}
\caption{Growth rates of unstable modes in quark matter of net quark density 0.7 fm$^{-3}$ and temperature 45 MeV without and with the collisional term using a light quark scattering cross section that has a value of 3 mb and is isotropic.}
\label{dsp_cll_n_07_T_045}
\end{figure}

\begin{figure}[htbp]
\vspace{-0cm}
\includegraphics[width=0.4 \textwidth]{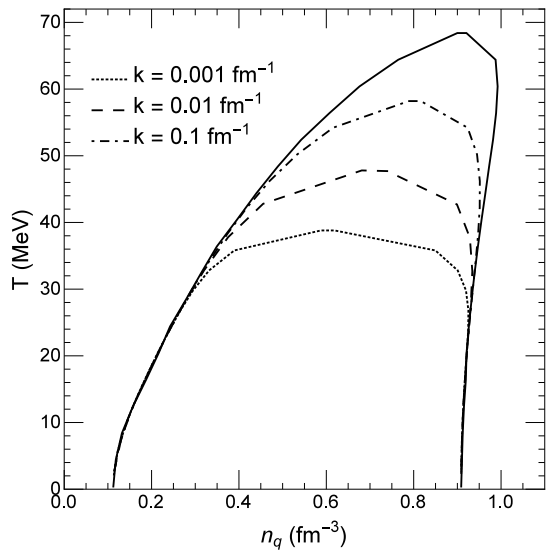}
\caption{The spinodal region calculated with the inclusion of the collisional term using an isotropic quark scattering cross section of 3 mb for different values of the wave number of unstable mode.}
\label{region_cll}
\end{figure}

In Fig.~\ref{dsp_cll_n_07_T_045}, we show by dashed and solid lines the dispersion relation or growth rate of the unstable mode in a quark matter of net quark density 0.7 fm$^{-3}$ and temperature 45 MeV with and without the collisional effect using
an isotropic light quark scattering cross section of 3 mb. The collisional effect is seen to reduce $\Gamma_k$ by almost a constant value for all wave number $k$.  Although the reduction is small, its effect is important for soft unstable modes. In this particular example, unstable modes with $k<0.005~\mathrm{fm}^{-1}$ disappears after the inclusion of collisional effects. As a result, the spinodal instability regions for unstable modes of longer wavelength  become smaller than those of shorter wavelength  as shown in Fig.~\ref{region_cll} by dotted, dashed, and dash-dotted lines for the spinodal boundaries of unstable modes of different wave numbers of 0.001, 0.01, and 0.1 fm$^{-1}$, respectively. Compared to the entire spinodal instability region shown by the solid curve, the area of the spinodal region in the temperature and density plane shrinks as the wave number of the unstable mode $k$ decreases. This behavior is opposite to that shown in the previous Section, where the high $k$ modes disappear due to quantum effects, and the spinodal region shrinks as $k$ increases. We note that in the absence of collisions, the spinodal boundary is independent of the wave number of unstable modes in the linearized Boltzmann approach. 

\section{Summary}

Using the linear response theory based on the PNJL model as well as its semiclassical approximation, the linearized Boltzmann equation based on the NJL model, we have studied the mean-field and confinement as well as collisional effects on the spinodal instabilities in a baryon-rich quark matter. In particular, we have studied the spinodal boundaries of unstable modes of different wavelengths in the temperature and density plane as well as their dispersion relations or growth rates.  We have found that in the long wavelength limit, the spinodal boundaries obtained in our study coincides with those determined from the isothermal spinodal instability in the thermodynamical approach.  Also, the vector interaction is found to suppress the spinodal instabilities of unstable modes of all wavelengths as a result of its repulsive effect in baryon-rich quark matter.  Due to the confinement effect, the Polyakov loop in the PNJL model is found to enhance the spinodal instablities of a quark matter compared to those from the NJL model.  We have further found that while the collisional effect reduces the growth rate of all unstable modes, the quantum effect further suppresses unstable modes of shorter wavelength.  For the PNJL model, we have found that the typical growth rate of unstable modes is only about 0.01 fm$^{-1}$, corresponding to a time duration of about 100 fm/$c$ for the instability or density fluctuation to grow.  Since this time is much longer than the lifetime of QGP produced in a heavy ion collisions, it is thus of great interest to study how unstable modes would grow if one goes beyond the linear response or small amplitude limit as studied here. In this respect, the transport equation discussed in Section IV provides a convenient framework to address this question either by studying the time evolution of density fluctuations in a confined or an expanding quark matter.   
Such a study is underway and will be reported in a future publication. 

\section*{Acknowledgements}

We thank Yunpeng Liu for helpful discussions. This work was supported in part by the Welch Foundation under Grant No. A-1358.

\bibliography{references}

\end{document}